 \documentclass[final,5p,times,twocolumn,number,sort&compress]{elsarticle}

 \usepackage{epsfig}

\usepackage{amssymb}
\usepackage{lipsum}
\usepackage{amsthm}

\journal{Physics Letters A}

\usepackage{mathtools,leftindex,tensor,mhchem}
\usepackage{booktabs}
\usepackage{siunitx}
\usepackage{enumitem}
\usepackage{dcolumn}
\usepackage{mathrsfs}
\usepackage{comment}
\usepackage{hyperref}
\usepackage{amsmath,accents}
\usepackage{mathtools}
\usepackage{bbm}
\usepackage{bm}
\usepackage{amsfonts}
\usepackage{amssymb}
\usepackage{amsthm}
\usepackage{mathrsfs}
\usepackage[scr=rsfs,cal=boondox]{mathalfa}
\usepackage{enumerate}
\usepackage{graphicx}
\usepackage{xypic}
\usepackage{filecontents}
\usepackage[dvipsnames]{xcolor}
\usepackage[scr=rsfs,cal=boondox]{mathalfa}

\usepackage{subcaption}
\usepackage{dblfloatfix}

\usepackage{lipsum}

\usepackage{stackengine}
\stackMath
\newcommand\tsup[2][2]{%
 \def\useanchorwidth{T}%
  \ifnum#1>1%
    \stackon[-1.3ex]{\tsup[\numexpr#1-1\relax]{#2}}{\mathchar"307E}%
  \else%
    \stackon[-1ex]{#2}{\mathchar"307E}%
  \fi%
}
\hypersetup{
    bookmarks=true,         
    pdftoolbar=true,        
    pdfmenubar=true,        
    pdffitwindow=true,     
    pdfstartview={FitH},     
    pdftitle={My title},     
    pdfauthor={author},      
    pdfsubject={Subject},    
    pdfcreator={Creator},    
    pdfproducer={Producer},  
    pdfkeywords={keyword1} 
    pdfnewwindow=true,       
    colorlinks=true ,        
    linkcolor=Blue  ,        
    citecolor=Blue,      
    filecolor=Blue,       
    urlcolor=Blue            
}

\usepackage{scalerel}
\usepackage{tikz}
\usetikzlibrary{svg.path}

\definecolor{orcidlogocol}{HTML}{A6CE39}
\tikzset{
  orcidlogo/.pic={
    \fill[orcidlogocol] svg{M256,128c0,70.7-57.3,128-128,128C57.3,256,0,198.7,0,128C0,57.3,57.3,0,128,0C198.7,0,256,57.3,256,128z};
    \fill[white] svg{M86.3,186.2H70.9V79.1h15.4v48.4V186.2z}
                 svg{M108.9,79.1h41.6c39.6,0,57,28.3,57,53.6c0,27.5-21.5,53.6-56.8,53.6h-41.8V79.1z M124.3,172.4h24.5c34.9,0,42.9-26.5,42.9-39.7c0-21.5-13.7-39.7-43.7-39.7h-23.7V172.4z}
                 svg{M88.7,56.8c0,5.5-4.5,10.1-10.1,10.1c-5.6,0-10.1-4.6-10.1-10.1c0-5.6,4.5-10.1,10.1-10.1C84.2,46.7,88.7,51.3,88.7,56.8z};
  }
}

\newcommand\orcidicon[1]{\href{https://orcid.org/#1}{\mbox{\scalerel*{
\begin{tikzpicture}[yscale=-1,transform shape]
\pic{orcidlogo};
\end{tikzpicture}
}{|}}}}

\begin{document}

\begin{frontmatter}
    \title{Charged particle bound orbits around magnetized Schwarzschild black holes: S2 star and hotspot applications
    }
    
\author[mainaddress1,mainaddress2,new]
{Uktamjon Uktamov}
\ead{uktam.uktamov11@gmail.com}%

\author[mainaddress3]{Mohsen Fathi\orcidicon{0000-0002-1602-0722}}
\ead{mohsen.fathi@ucentral.cl}

\author[mainaddress5,mainaddress4,urdu]{Javlon Rayimbaev}
\ead{javlon@astrin.uz}

\address[mainaddress1]{School of Physics, Harbin Institute of Technology, Harbin 150001, People’s Republic of China}

\address[mainaddress2]{Tashkent University of Applied Sciences, Gavhar Str. 1, Tashkent 100149, Uzbekistan}

\address[new]{Institute for Advanced Studies, New Uzbekistan University,
Movarounnahr str. 1, Tashkent 100000, Uzbekistan}

\address[mainaddress3]{Centro de Investigaci\'{o}n en Ciencias del Espacio y F\'{i}sica Te\'{o}rica (CICEF), Universidad Central de Chile, La Serena 1710164, Chile}

\address[mainaddress5]{National University of Uzbekistan, Tashkent 100174, Uzbekistan}

\address[mainaddress4]{Tashkent State Technical University, Tashkent 100095, Uzbekistan}

\address[urdu]{Urgench State University, Kh. Alimjan Str. 14, Urgench 221100, Uzbekistan}

\date{Received: date / Accepted: date}

\begin{abstract}
The dynamics of charged particles around magnetized black holes provide valuable insights into astrophysical processes near compact objects. In this work, we investigate the bound and unbound trajectories of charged particles in the vicinity of a Schwarzschild black hole immersed in an external, uniform magnetic field. By analyzing the effective potential and solving the corresponding equations of motion, we classify the possible orbital configurations and identify the critical parameters governing the transition between stable and escape trajectories. The influence of the magnetic field strength and particle charge on the orbital structure, energy, and angular momentum is systematically explored. Applications of the obtained results are discussed in the context of the S2 star orbiting Sagittarius A* and the motion of bright hotspots detected near the event horizon, offering a potential interpretation of recent observations in terms of magnetized dynamics. The study contributes to a deeper understanding of charged-particle motion around black holes and its relevance to high-energy astrophysical phenomena in the galactic center. Finally, we test our model by fitting it to real data from the observed trajectory of the S2 star using a statistical Markov Chain Monte Carlo (MCMC) method. This allows us to find the best estimates for magnetic field and charge of the S2 star.
\end{abstract}


\end{frontmatter}

\section{Introduction \label{Sec:introduction}}

The motion of charged particles in the vicinity of black holes provides a powerful means of probing the interplay between gravitation, electromagnetism, and relativistic dynamics. The presence of external magnetic fields near compact objects can substantially modify particle trajectories, giving rise to rich orbital structures with direct astrophysical implications \cite{FrolovShoom2010,Lim2015,TursunovStuchlikKolos2016,NarzilloevAbdujabbarovBambiAhmedov2019,Abdulxamidov2023,Ladino2023,Uktamov:2024ckf,2024PhRvD.110h4084U,Oteev2025}. Such magnetized environments are expected to occur naturally in accretion flows and jet-launching regions surrounding supermassive black holes, where strong magnetic fields influence both plasma motion and radiation processes \cite{McKinney2015,Sarkar2016,Ripperda2022,chakraborty_gravitational_2022,Janiuk2022,Crinquand2022,chakraborty_magnetic_2024,chakraborty_black_2024,Nathanail2025,Chatterjee2025}.

Among the most straightforward yet most instructive models for investigating these effects is the Schwarzschild black hole (SBH) immersed in a uniform magnetic field, often described through Wald’s formalism \cite{Wald1974,Galtsov1978,Tursunov2013,Kolos2015,RAYIMBAEV2024101516,Rayimbaev2023}. In this configuration, the magnetic field leaves the background geometry unchanged but modifies charged-particle motion via the Lorentz force. As a result, the corresponding trajectories exhibit intricate dependencies on the particle’s charge, energy, angular momentum, and the magnetic field strength. Analyses of such systems not only yield theoretical insights but also offer possible explanations for observed phenomena near black holes, such as quasi-periodic oscillations, flares, and hotspot orbits detected around Sagittarius~A* (Sgr~A*) \cite{Stuchlik2013,Lin2023,Aimar2023,RAHMATOV2024143,Levis2024}.

Recent advances in high-resolution infrared interferometry, particularly through the GRAVITY Collaboration, have enabled precise tracking of the S2 star orbiting Sgr~A* and the detection of compact emission features close to the event horizon \cite{Abuter2018,abuter2018detection,Abuter2020,Abuter2021,DeMartino2021,Straub2023}. These observations open new possibilities for testing relativistic dynamics in strong-gravity regimes. In this context, observational constraints based on the S2 star and statistical analyses using MCMC techniques have recently been explored in different gravitational frameworks \cite{navarrete_testing_2026}. The study of charged particle motion in magnetized black hole spacetimes provides a sound theoretical framework for interpreting such data, offering constraints on the local magnetic field distribution, plasma properties, and the possible charge-to-mass ratios of radiating particles.

In this work, we analyze the bound and unbound trajectories of charged particles in a Schwarzschild spacetime embedded in an external, asymptotically uniform magnetic field. Using the Hamilton–Jacobi formalism, we derive the equations of motion and examine the structure of the corresponding effective potential, identifying the regions of stable, unstable, and escape orbits. The influence of the magnetic-field parameter and the particle’s charge on the critical radii of circular motion, including the innermost stable circular orbit (ISCO), is explored in detail.

Furthermore, we apply the obtained results to two astrophysical contexts: (i) the orbital dynamics of the S2 star in the galactic center, and (ii) the motion of bright hotspots observed near Sgr~A*, whose periodic variability may reflect the effects of magnetized dynamics. We also employ the Levin–Perez-Giz (LP) taxonomy of relativistic orbits \cite{Levin2008,PerezGiz2008a,PerezGiz2009,Misra2010,Grossman2012,Gao2020,Lim2024} to classify bound trajectories in the magnetized Schwarzschild background.

The remainder of this paper is organized as follows. In Sect.~\ref{sec:motion}, we present the SBH spacetime immersed in a magnetic field and derive the equations of motion using the standard Lagrangian formalism, which are then analyzed for orbital stability. We subsequently constrain the magnetic parameter of the model in accordance with the observed orbital characteristics of the S2 star around Sgr~A*. In Sect.~\ref{sec:LP}, we explore the influence of different magnetic-field strengths on the bound motion of charged particles and provide numerical simulations of the corresponding trajectories. Finally, our conclusions are summarized in Sect.~\ref{sec:conclusion}.

In this work, we employ natural units ($G = c = 1$), adopt the metric signature $(-,+,+,+)$, and use primes to indicate derivatives with respect to the radial coordinate.

\section{Charged Particle Dynamics around SBHs in a Magnetic Field}\label{sec:motion}

We consider a SBH described by the line element
\begin{eqnarray}\label{eq:metric}
    ds^2=-f(r) dt^2 + f(r)^{-1} dr^2 + r^2 d\Omega^2,
\end{eqnarray}
in the standard Schwarzschild coordinates $x^\alpha=(t,r,\theta,\phi)$, where $d\Omega^2=d\theta^2+\sin^2\theta d\phi^2$. The lapse function is given by
\begin{equation}
    f(r) = 1 - \frac{2M}{r}.
\end{equation}
The black hole is immersed in an external, asymptotically uniform magnetic field of strength $B$. Following Wald's prescription \cite{Wald1974}, the electromagnetic four-potential is expressed as $A^\mu = \frac{B}{2}\xi^\mu_\phi$, with $\xi^\mu_\phi=(0,0,0,1)$ representing the azimuthal Killing vector. Consequently, the only non-zero covariant component of the potential is 
\begin{equation}
    A_\phi = \frac{1}{2} B r^2 \sin^2 \theta.
\end{equation}

\subsection{Equations of motion}

The motion of a charged particle with mass $m$ and electric charge $q$ is derived from the Lagrangian
\begin{eqnarray}\label{eq:Lagrangian}
    \mathcal{L} = \frac{1}{2} m g_{\mu\nu} u^\mu u^\nu + q A_\mu u^\mu,
\end{eqnarray}
and equivalently from the Hamilton-Jacobi equation \cite{Uktamov:2024ckf}:
\begin{eqnarray}\label{eq:HamiltonJacobi}
    g^{\mu\nu} \left( \frac{\partial S}{\partial x^\mu} - q A_\mu \right) 
    \left( \frac{\partial S}{\partial x^\nu} - q A_\nu \right) = - m^2,
\end{eqnarray}
where the action is $S = -Et + L\phi + S_\theta(\theta) + S_r(r)$. Confining motion to the equatorial plane ($\theta = \pi/2$), the equations of motion reduce to
\begin{subequations}\label{eq:motion}
\begin{align}
    & \dot{t} = \frac{\mathcal{E}}{f(r)},\\
    & \dot{\phi} = \frac{l}{r^2} - \beta,\\
    & \dot{r}^2 = \mathcal{E}^2 - V_\text{eff}(r),
\end{align}
\end{subequations}
where $\mathcal{E} = E/m$ is the specific energy, $l = L/m$ is the specific angular momentum, and $\beta = q B / (2 m)$ is the magnetic parameter. Hence, the radial motion as a function of $\phi$ is governed by
\begin{equation}
    \left( \frac{dr}{d\phi} \right)^2 
   = \left( \frac{r^2}{l - \beta r^2} \right)^2 \Bigl[ \mathcal{E}^2 - V_{\rm{eff}}(r) \Bigr] \coloneqq  P(r),
   \label{eq:drdphi}
\end{equation}
in which the effective potential is
\begin{eqnarray}\label{eq:Veff}
    V_\text{eff}(r) = f(r) \left[ 1 + \left( \frac{l}{r} - \beta r \right)^2 \right].
\end{eqnarray}

\subsection{Circular orbits and stability}

\begin{figure*}[ht!]
\centering
\includegraphics[width=0.28\textwidth]{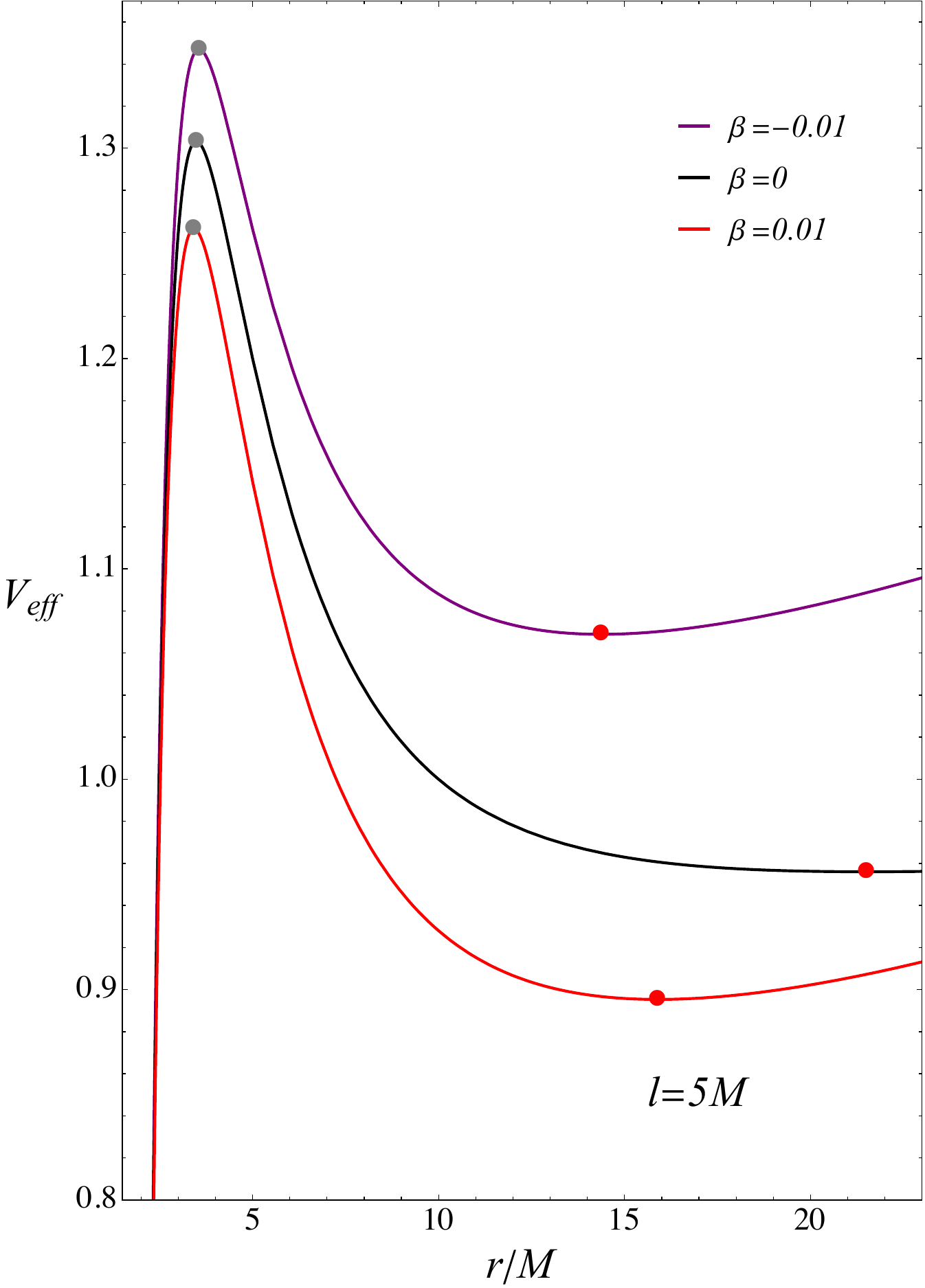} (a)\qquad
\includegraphics[width=0.28\textwidth]{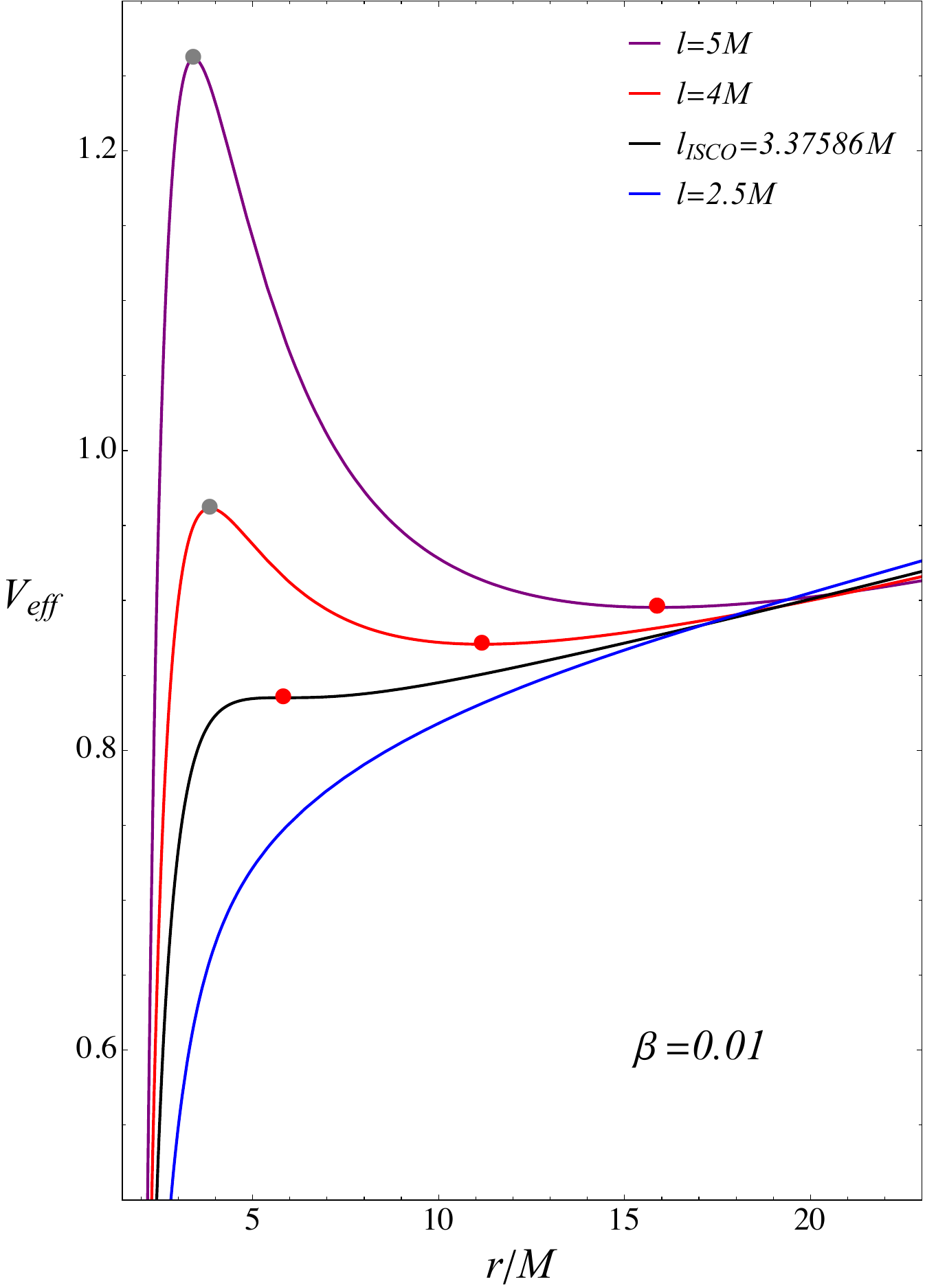} (b)
\caption{Effective potential $V_\text{eff}(r)$ for charged particles near a SBH: (a) varying $\beta$ with fixed $l$, and (b) varying $l$ with fixed $\beta$. Red points correspond to stable circular orbits, gray points to unstable orbits.}
\label{effective}
\end{figure*}

Figure \ref{effective} shows the effective potential $V_\text{eff}(r)$ as a function of $r$. 
Minima correspond to stable circular orbits (red points), and maxima correspond to unstable orbits (gray points). For $\beta>0$, the Lorentz force is repulsive, reducing the radius of unstable orbits, while $\beta<0$ increases it due to attraction. 

\begin{figure*}[ht!]
\centering
\includegraphics[width=0.35\textwidth]{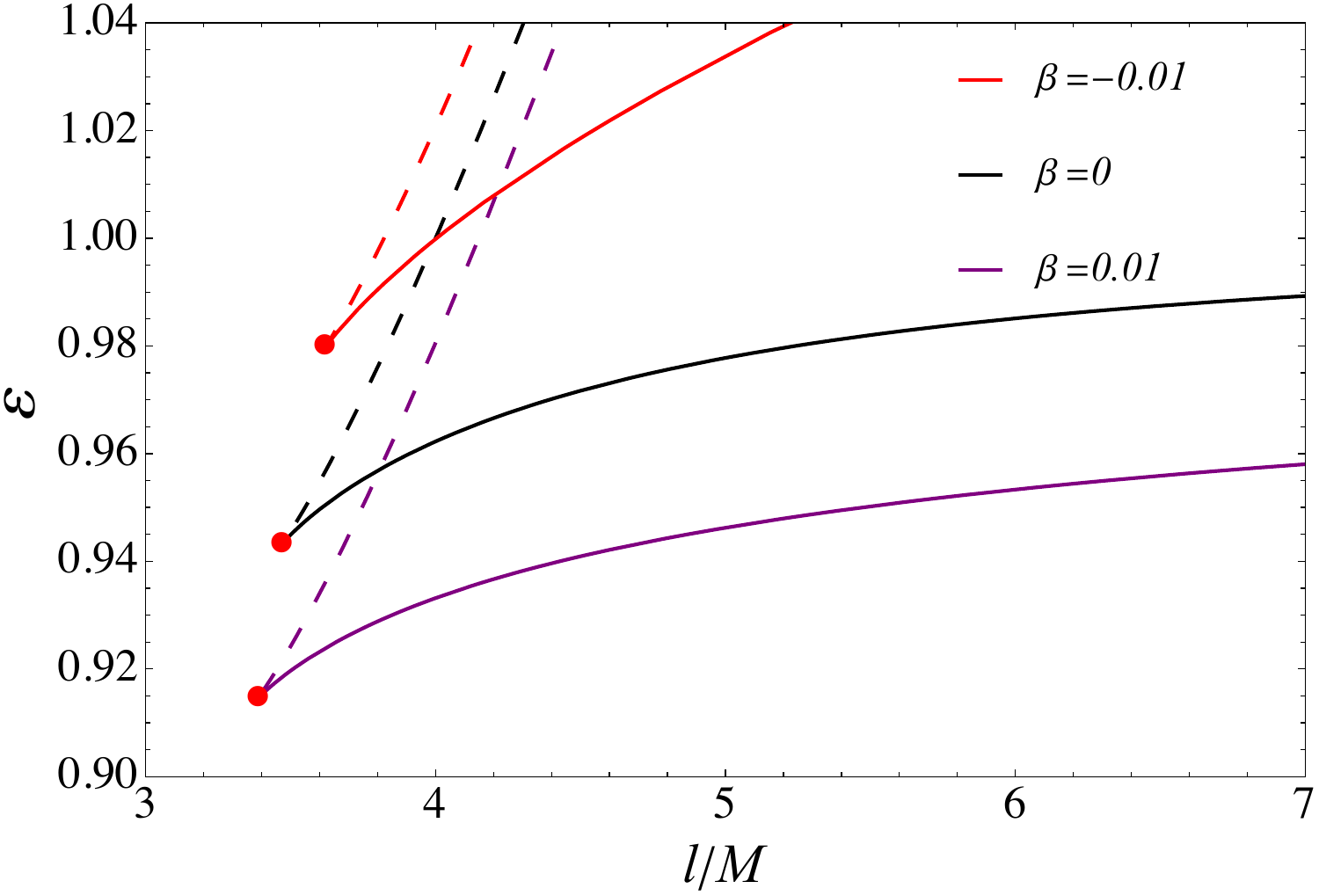}
\caption{Specific energy $\mathcal{E}_c$ versus angular momentum $l_c$ for circular orbits. Solid lines indicate stable orbits, dashed lines indicate unstable orbits. Red points mark the ISCO (see also Table \ref{Table1}). }
\label{fig.l_e_space}
\end{figure*}

In general, the specific angular momentum $l$ and specific energy $\mathcal{E}$ for circular orbits of radius $r_c$ are obtained from $\mathcal{E}_c^2 = V_\text{eff}(r_c)$ and $V'_\text{eff}(r_c) = 0$:
\begin{align}
l_c &= \frac{r_c \Big[ \sqrt{\beta^2 r_c^4 f(r_c)^2 + M (r_c - 3 M)} 
        - \beta M r_c \Big]}{r_c - 3 M}, \label{eq:lc} \\[2mm]
\mathcal{E}_c^2 &=  f(r_c) \Bigg[ 1 
        + \frac{\biggl( \sqrt{\beta^2 r_c^4 f(r_c)^2 + M (r_c - 3 M)} 
        - \beta r_c^2 f(r_c) \biggr)^2}{(r_c - 3 M)^2} \Bigg]. \label{eq:Ec}
\end{align}
In Fig.~\ref{fig.l_e_space}, we show the relationship between $\mathcal{E}_c$ and $l_c$ for different values of the magnetic parameter $\beta$.

Moreover, the innermost stable circular orbit (ISCO) is defined by $V''_\text{eff} = 0$, with stable orbits satisfying $V''_\text{eff} > 0$ and unstable ones satisfying $ V''_\text{eff} < 0$. Table \ref{Table1} lists the critical radii for three values of $\beta$, showing the impact of this parameter on the critical radius.
\begin{table}[ht!]
    \centering
    \begin{tabular}{|c|c|}
     \hline
       $\beta$ & $r_\text{ISCO}$  \\
    \hline
    $-0.01$   & $5.85955 M$  \\
      $0.0$   & $6 M$ \\
      $0.01$  & $5.8414 M$ \\
   \hline
    \end{tabular}
    \caption{Critical radii of circular orbits for different values of the magnetic parameter $\beta$.}
    \label{Table1}
\end{table}
\begin{figure*}[ht!]
\centering
\includegraphics[width=0.3\textwidth]{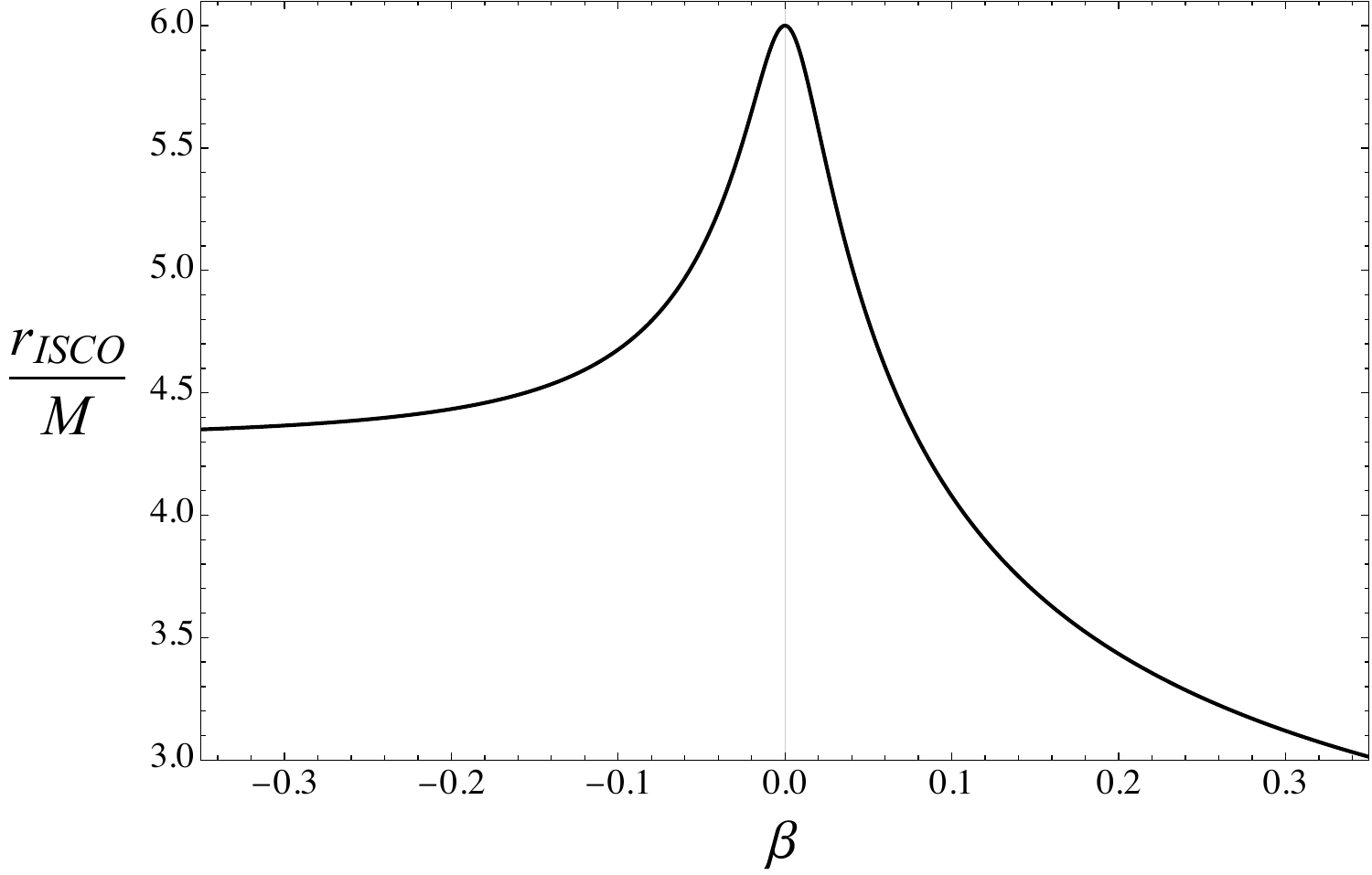} (a)\quad
\includegraphics[width=0.3\textwidth]{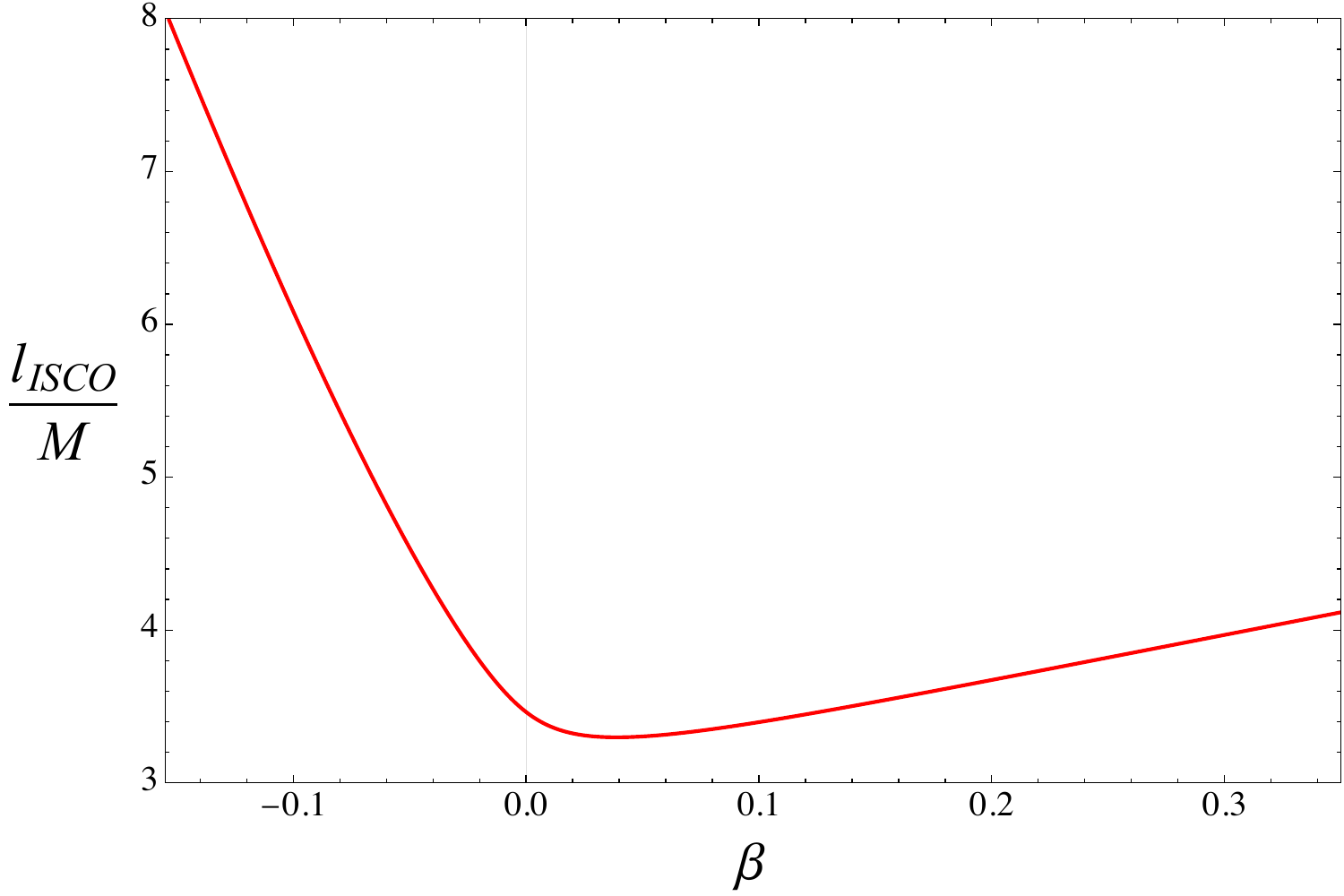} (b) \quad
\includegraphics[width=0.3\textwidth]{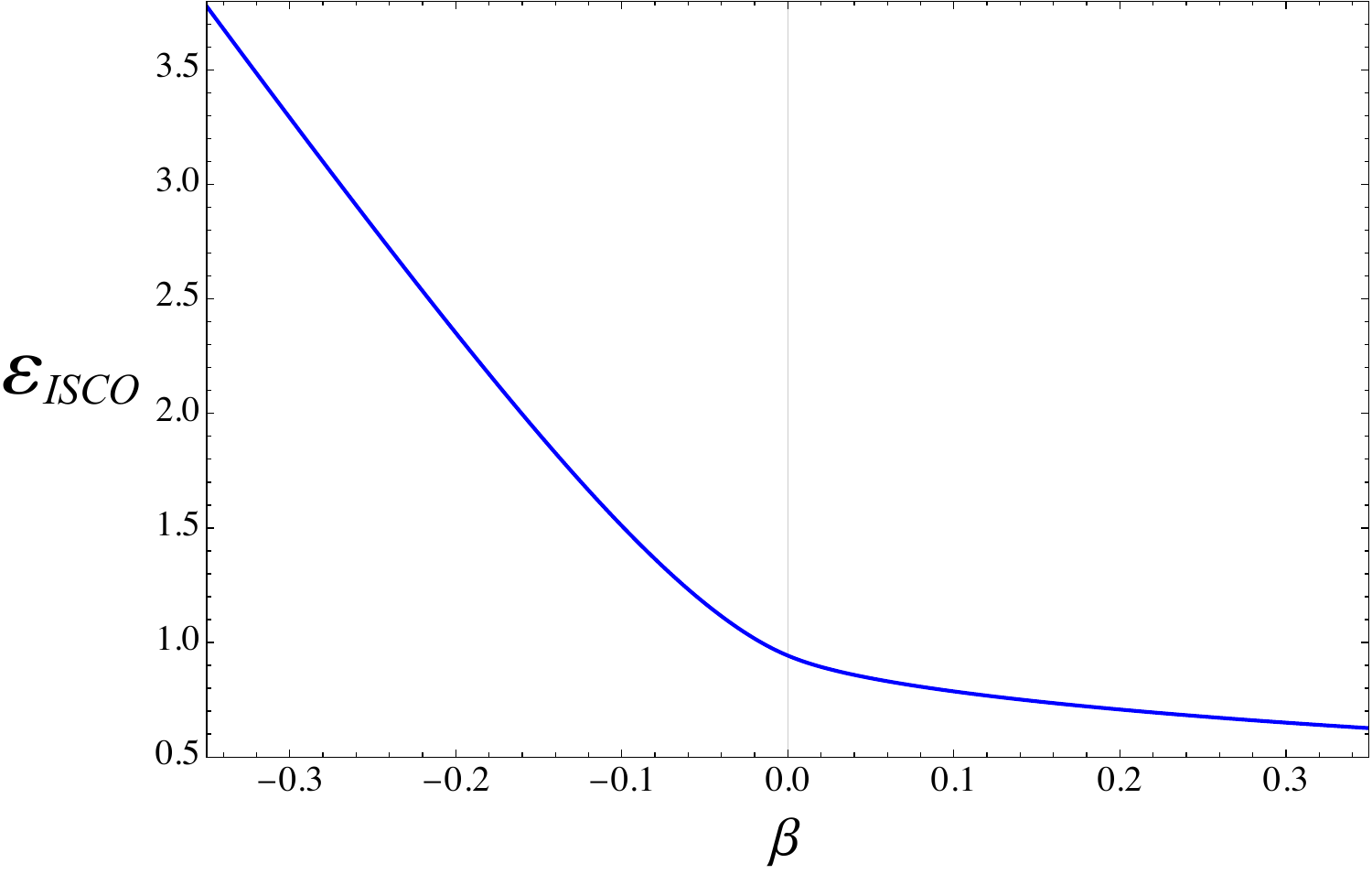} (c)
\caption{Dependence of ISCO parameters on $\beta$: (a) radius $r_\text{ISCO}$, (b) angular momentum $l_\text{ISCO}$, and (c) energy $\mathcal{E}_\text{ISCO}$.}
\label{fig.ISCO}
\end{figure*}

Figure~\ref{fig.ISCO} illustrates the dependence of the critical radius on $\beta$, along with the corresponding constants of motion at the ISCO.

\subsection{Applications to hotspot orbits around Sgr A*}

\begin{figure*}[ht!]
\centering
\includegraphics[width=0.45\textwidth]{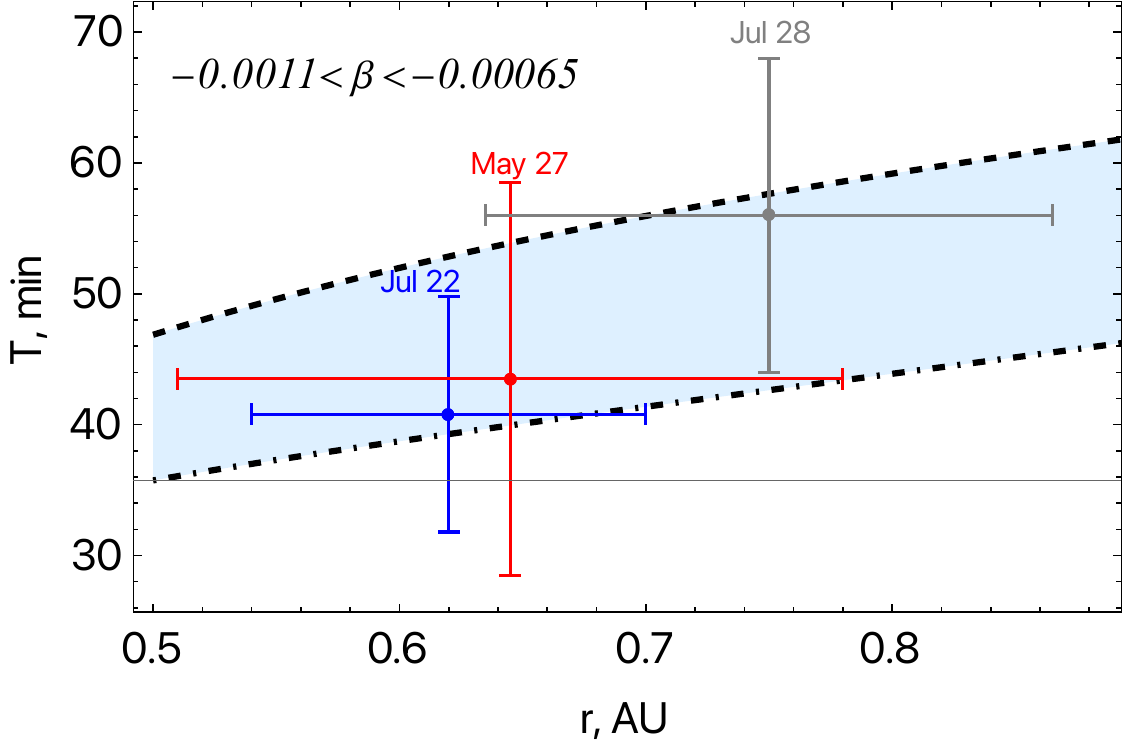}
\caption{Orbital period $T$ versus radius for various magnetic parameters $\beta$.}
\label{Period}
\end{figure*}

Using GRAVITY data \cite{abuter2018detection}, the orbital period $T$ of a charged particle on a circular orbit is
\begin{equation}
    T = \frac{2 \pi}{\Omega_K}, 
\label{eq:T}
\end{equation}
where
\begin{equation}
    \Omega_K = \frac{\dot{\phi}}{\dot{t}} = \frac{f(r) \left( l_c - \beta r^2 \right)}{r^2 \mathcal{E}_c},
    \label{eq:OmegaK}
\end{equation}
with $l_c$ and $\mathcal{E}_c$ given by Eqs. \eqref{eq:lc} and \eqref{eq:Ec}. Figure \ref{Period} shows the period as a function of orbital radius for different $\beta$, constraining $-0.0011 < \beta < -0.00065$ for Sgr A* with $M = 4.14 \times 10^6 M_\odot$.

\section{Relativistic bound orbits and LP taxonomy}\label{sec:LP}

To analyze the bound motion of charged particles, we employ the LP taxonomy \cite{Levin2008}, which allows us to determine the specific energy $\mathcal{E}$ and specific angular momentum $l$ for a given orbit. From Eq.~\eqref{eq:drdphi}, bound orbits satisfy $P(r) = 0$, whose two roots can be expressed as
\begin{equation}
    r_1 = \frac{\lambda}{1 - e}, \quad r_2 = \frac{\lambda}{1 + e}, 
\end{equation}
where $\lambda$ and $e$ are the latus rectum and eccentricity of the orbit, respectively \cite{chandrasekhar1998mathematical}.  

Using Eq.~\eqref{eq:drdphi}, the specific angular momentum and energy of a charged particle corresponding to a given $(\lambda,e)$ are obtained as
\begin{align}
    l &= \frac{\mathcal{A} - \mathcal{B}}{\Psi}, \label{eq:angular}\\
    \mathcal{E} &= \frac{1}{\Psi}{\sqrt{\frac{\mathcal{C}\left(\mathcal{D} + \mathcal{G} + \mathcal{H} - \lambda\right)}{\Xi}}}, \label{eq:energy}
\end{align}
where the definitions of $\mathcal{A}$, $\mathcal{B}$, $\mathcal{C}$, $\mathcal{D}$, $\mathcal{G}$, $\mathcal{H}$, $\Psi$, and $\Xi$ are provided in \ref{app:A}.  

Assuming that the particle begins at $r = r_1$ and $\phi = 0$, the evolution of the azimuthal angle over one period is
\begin{equation}\label{eq:q_parameter}
    q + 1 = \frac{\Delta \phi}{2 \pi} = \frac{1}{\pi} \int_{r_1}^{r_2} \frac{dr}{\sqrt{P(r)}},
\end{equation}
where $q = \omega + {v}/{z}$ characterizes the orbit shape, with $\omega$, $v$, and $z$ representing the whirl, vertex, and zoom numbers. The latus rectum $\lambda$ for a chosen trajectory is obtained by numerically integrating Eq.~\eqref{eq:q_parameter} and applying the bisection method to determine its precise value.

Table~\ref{Table2} provides representative numerical values of the latus rectum $\lambda$, specific angular momentum $l$, and specific energy $\mathcal{E}$ for selected eccentricity $e$ and magnetic parameter $\beta$. These parameters correspond to the periodic orbits selected under the LP classification.
\begin{table}[ht!]
    \centering
    \begin{tabular}{|c|c|c|c|c|c|}
     \hline
       $(z,\omega,v)$ & $\beta$ & $e$ & $\lambda$ & $l$ & $\mathcal{E}$ \\
    \hline
      (1,0,1) & -0.01 & 0.8 & $7.57692 M$ & $4.53508 M$ & 1.08749 \\
      (1,0,1) & 0.0   & 0.8 & $8.56483 M$ & $3.85943 M$ & 0.98033 \\
      (1,0,1) & 0.001 & 0.8 & $8.27243 M$ & $3.83679 M$ & 0.97681 \\
      (2,0,1) & -0.01 & 0.8 & $9.43026 M$ & $5.14769 M$ & 1.13158 \\
      (2,0,1) & 0.0   & 0.8 & $11.145 M$  & $4.06822 M$ & 0.98450 \\
      (2,0,1) & 0.001 & 0.8 & $9.69877 M$ & $3.93809 M$ & 0.97969 \\
      (3,0,1) & -0.01 & 0.8 & $11.2205 M$ & $6.01062 M$ & 1.18105 \\
      (3,0,1) & 0.0   & 0.8 & $13.9918 M$ & $4.34877 M$ & 0.98751 \\
      (3,0,1) & 0.001 & 0.8 & $10.7349 M$ & $4.03252 M$ & 0.98142 \\
      (4,0,1) & -0.01 & 0.8 & $12.8577 M$ & $7.00283 M$ & 1.23147 \\
      (4,0,1) & 0.0   & 0.8 & $16.9124 M$ & $4.64228 M$ & 0.98959 \\
      (4,0,1) & 0.001 & 0.8 & $11.4448 M$ & $4.10277 M$ & 0.98247 \\
   \hline
    \end{tabular}
    \caption{Specific energy $\mathcal{E}$ and specific angular momentum $l$ for selected eccentricity $e$ and periodic orbits, with fixed values of the magnetic parameter $\beta$.}
    \label{Table2}
\end{table}
%

\subsection{Representative orbits}

\begin{figure*}[t]
\centering
\includegraphics[width=0.37\textwidth]{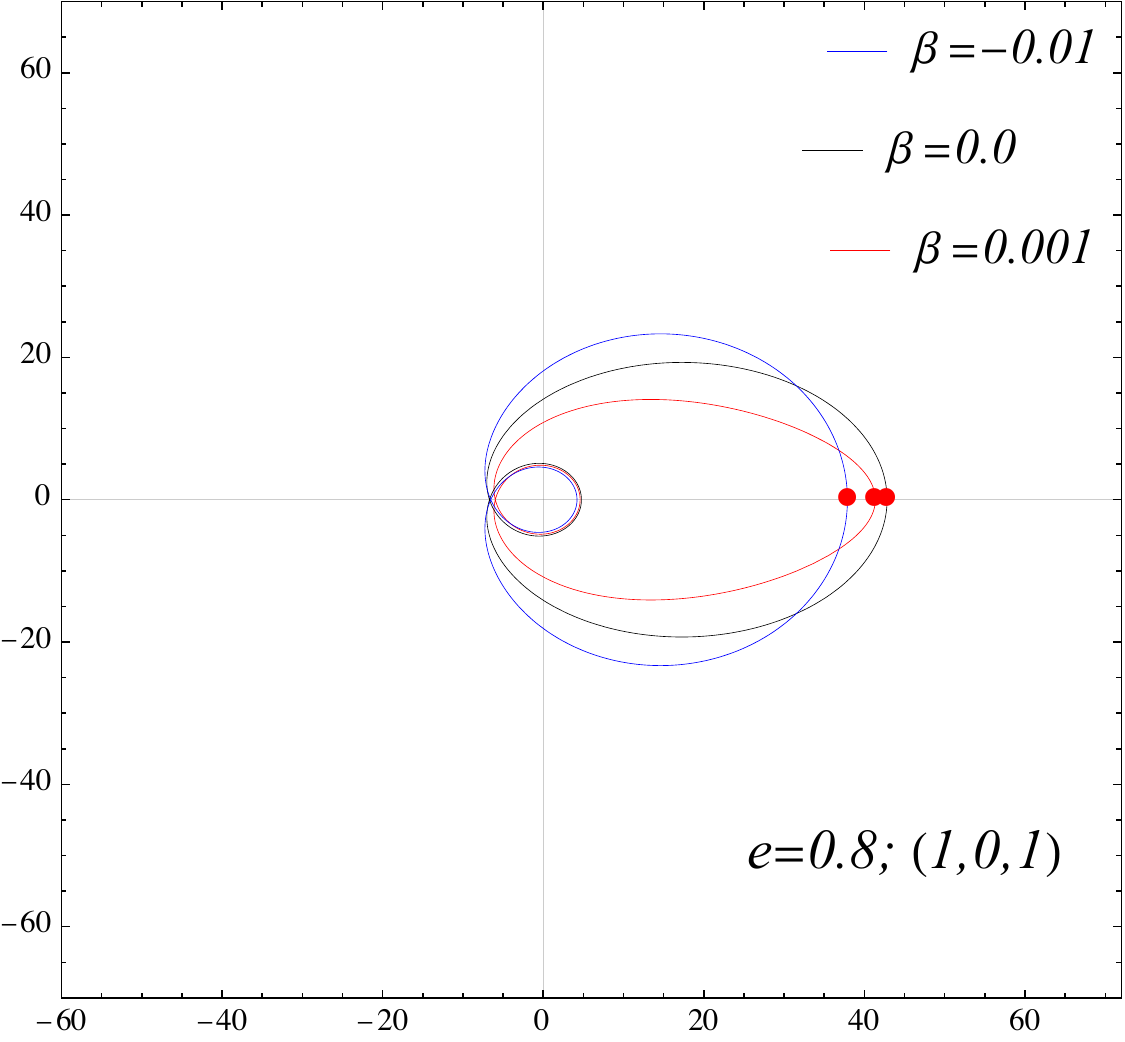} (a)\quad
\vspace{4mm}
\includegraphics[width=0.37\textwidth]{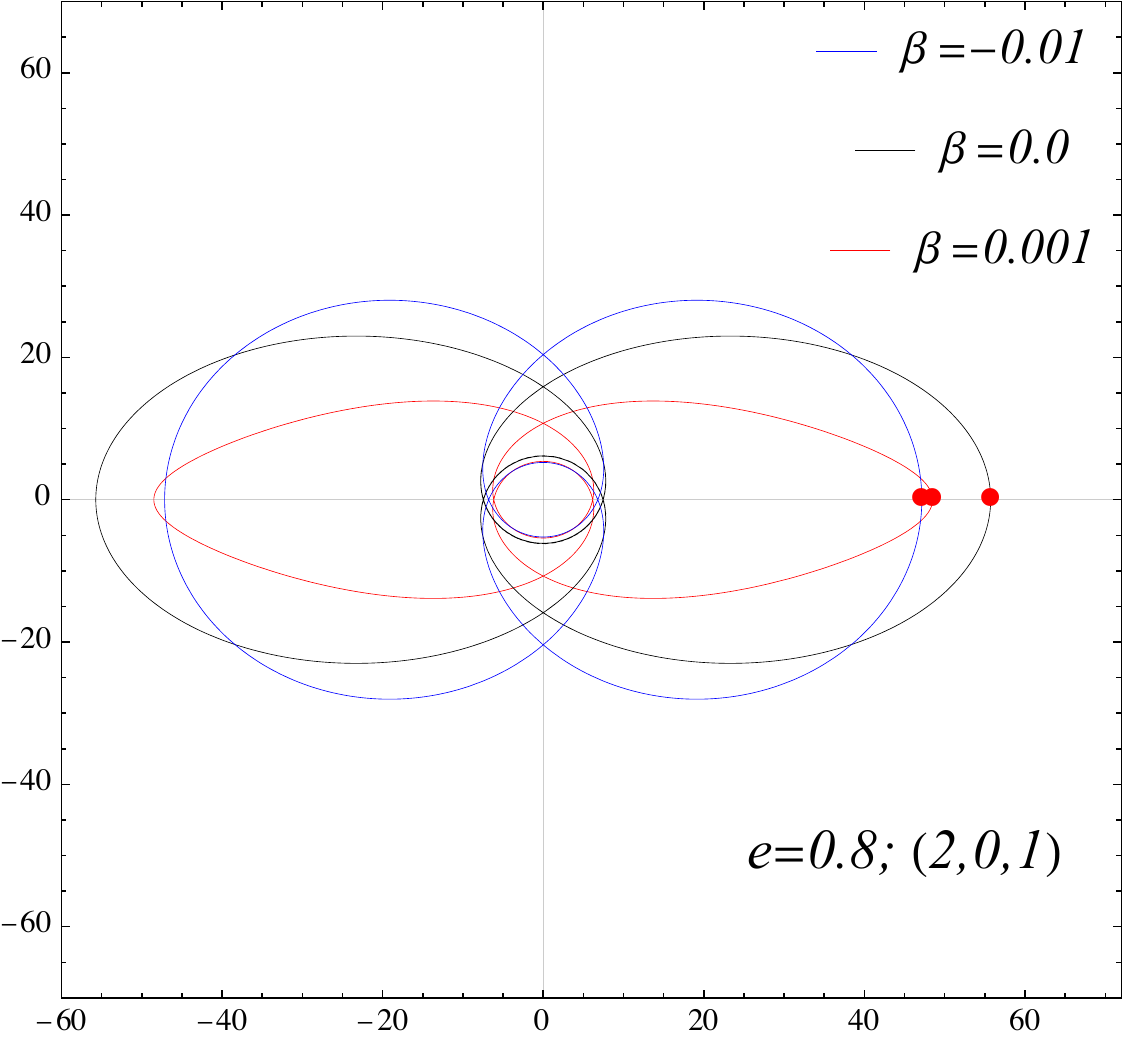} (b)
\vspace{4mm}
\includegraphics[width=0.37\textwidth]{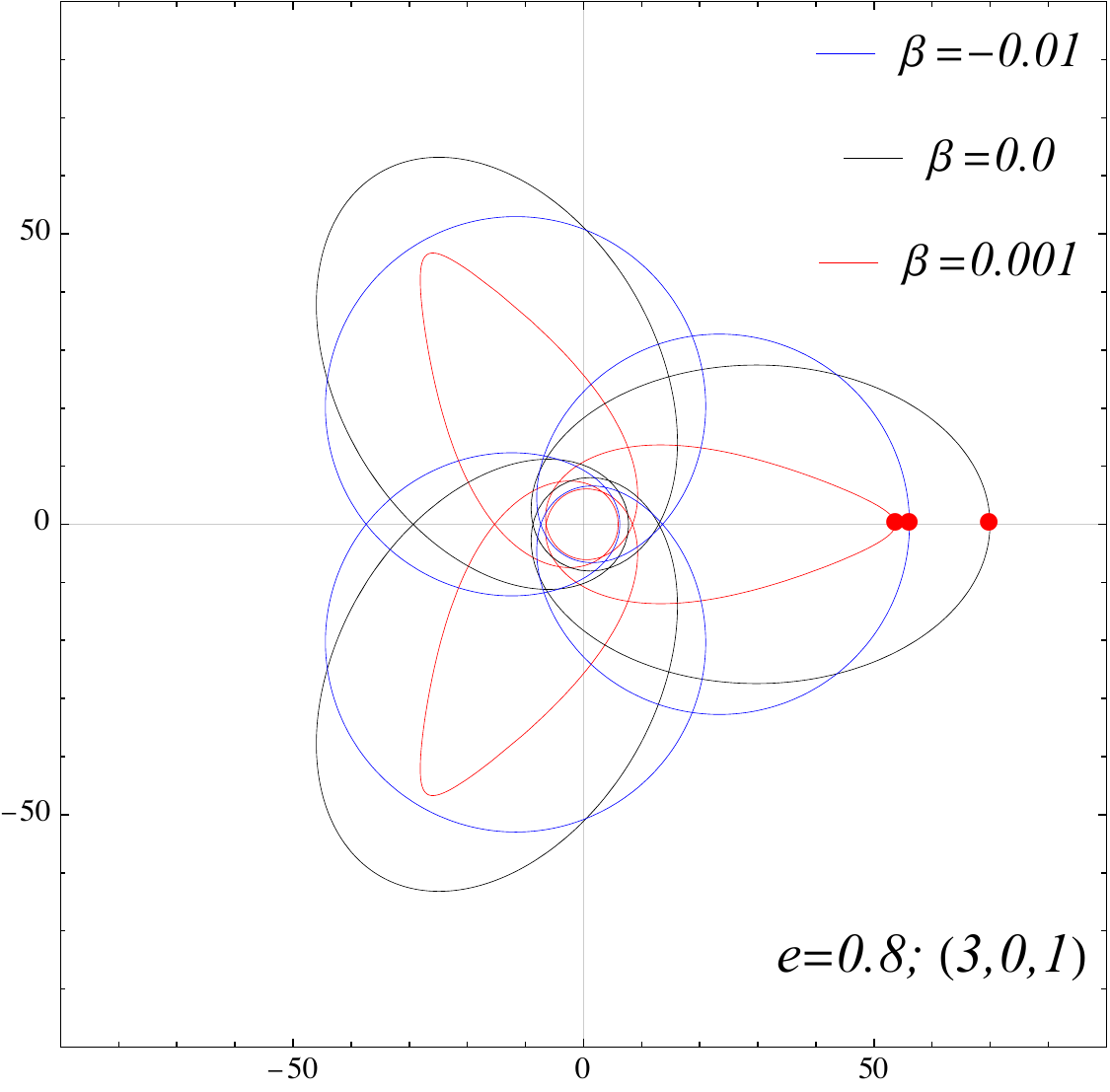} (c)\quad
\includegraphics[width=0.37\textwidth]{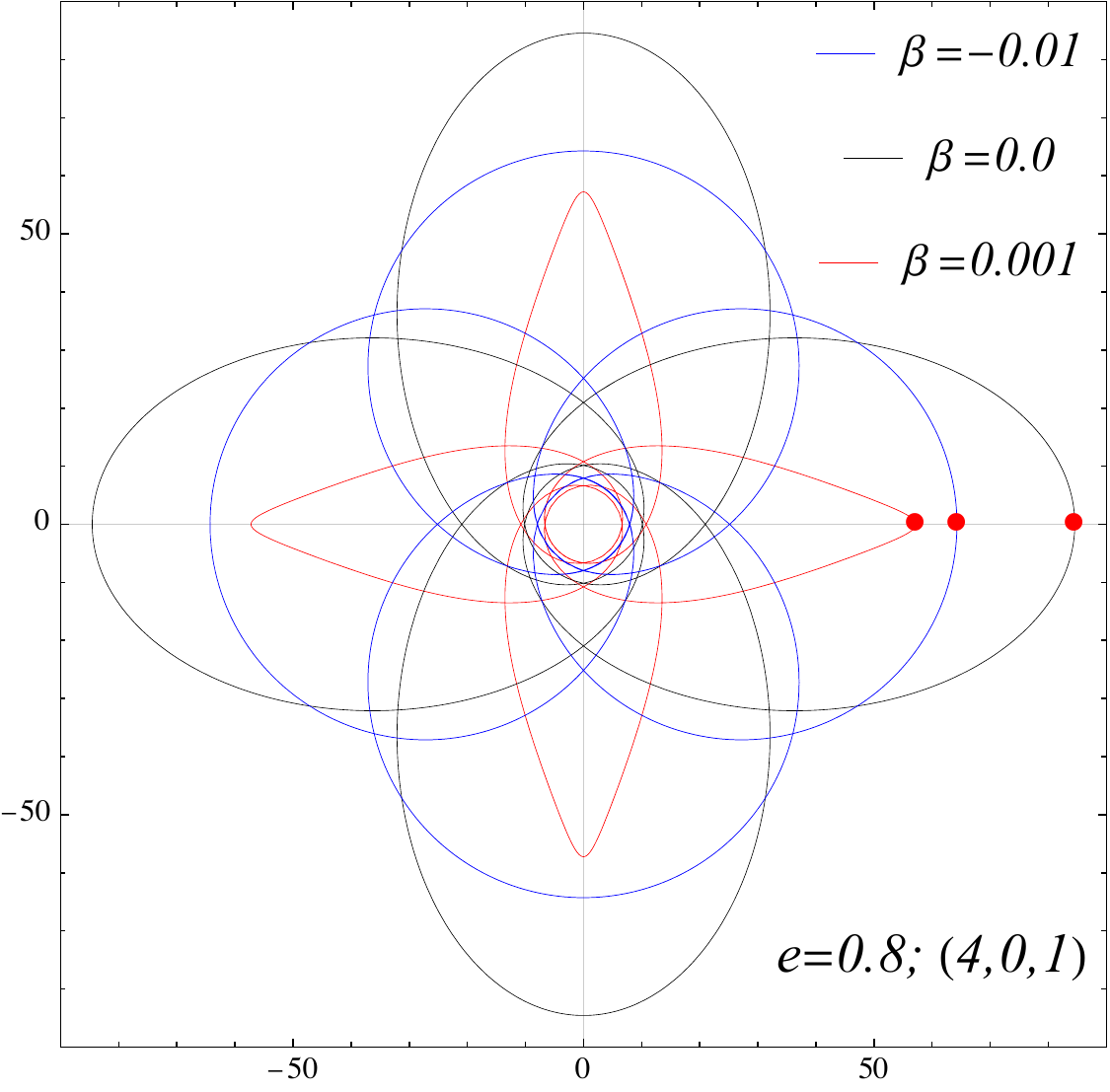} (d)
\caption{Bound orbits of charged particles for selected $(z,\omega,v)$ and fixed magnetic parameter $\beta$. The black hole is at the origin.}
\label{Orbits}
\end{figure*}

Figure~\ref{Orbits} shows the selected bound orbits for various $(z,\omega,v)$ and magnetic parameters $\beta$. The black hole is located at the origin, and the magnetic field shapes and extends the orbits.
%

\subsection{Estimating the magnetic parameter $\beta$ using the observed trajectory of the S2 star}

\begin{figure*}[ht!]
\centering
\includegraphics[width=0.5\textwidth]{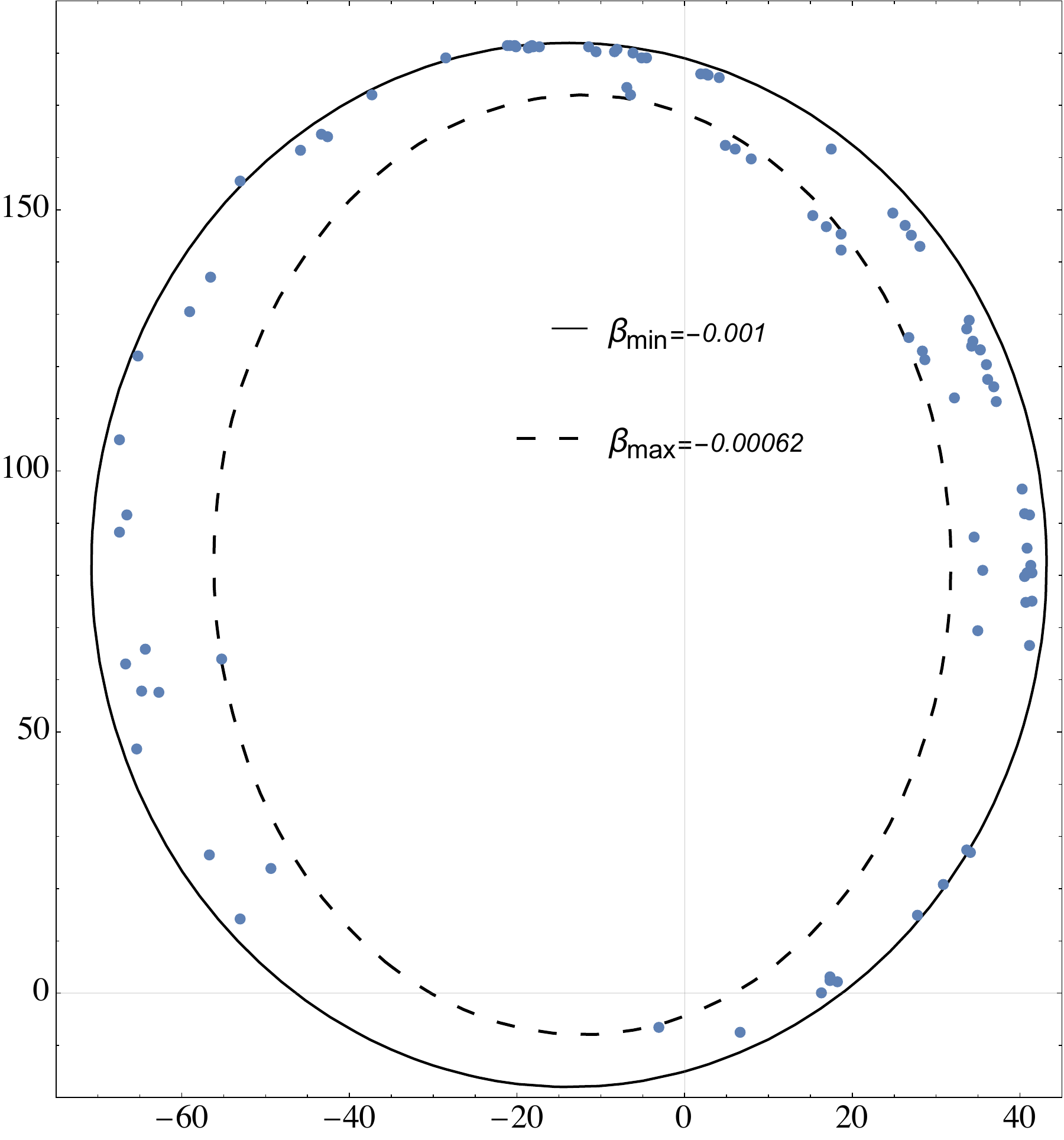}
\caption{Reconstructed trajectory of the S2 star around Sgr~A*, consistent with observational data \cite{2019Sci...365..664D}.}
\label{S2_trajectory}
\end{figure*}

By solving Eq.~(\ref{eq:q_parameter}) numerically for the orbital configuration defined by $(z, \omega, v) = (1, 0, 0)$, one obtains a nearly elliptical trajectory. Using the observational data of the S2 star's motion around Sgr~A* \cite{2019Sci...365..664D}, we can estimate the approximate value of the magnetic parameter $\beta$ associated with this orbit. Figure~\ref{S2_trajectory} displays the periodic trajectory that reproduces the observed motion of the S2 star. Our numerical analysis of Eq.~(\ref{eq:q_parameter}) based on the measured orbital parameters of S2 yields an estimated value of the magnetic parameter $-0.001 \lesssim\beta \lesssim -0.00062$, in which the trajectory is closer to the observational data. 
\begin{figure}[ht!]
\centering
\includegraphics[width=0.45\textwidth]{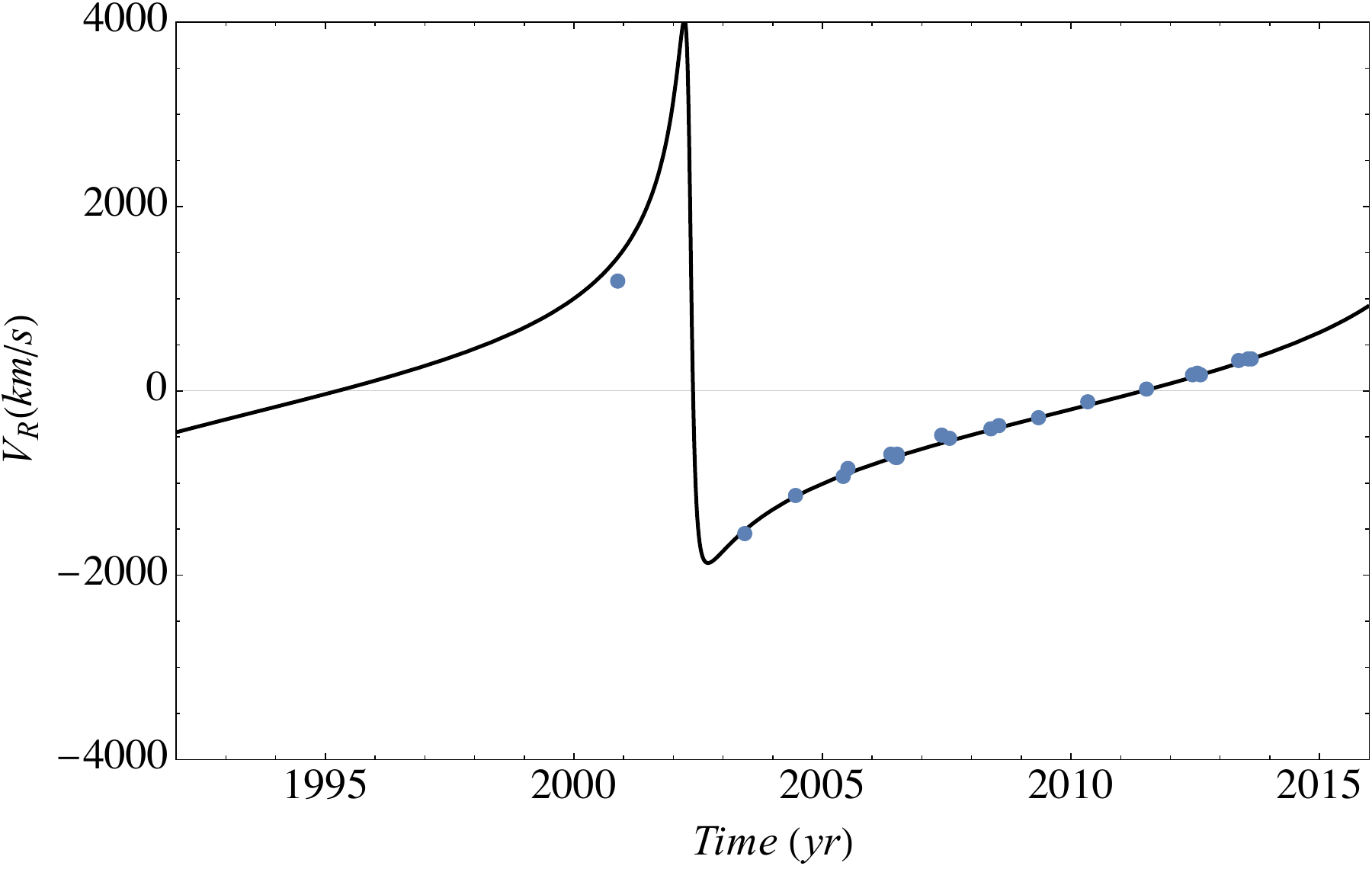}
\caption{Dataset of S2 star radial velocities used in our analysis.}
\label{fig.radial velocity}
\end{figure}

\section{Parametric constraints through MCMC analysis}\label{sec:MCMC}

We now proceed to infer the relevant model parameters via an MCMC exploration of the observational data (see, e.g., \cite{2025PDU....4902022U}).  
The total likelihood function adopted in our analysis is written as
\begin{eqnarray}\label{eq.Lik.}
    \mathcal{L} = \mathcal{L}_X + \mathcal{L}_Y + \mathcal{L}_{V}\,,
\end{eqnarray}
where the contributions associated with the astrometric coordinates and the radial velocity are respectively
\begin{subequations}\label{eq.Lik.2}
\begin{align}
    \mathcal{L}_X &= - \sum_i \left( \frac{X_i - \hat{X}_i}{\sigma_i^{x}} \right)^2 ,\\[2mm]
    \mathcal{L}_Y &= - \sum_i \left( \frac{Y_i - \hat{Y}_i}{\sigma_i^{y}} \right)^2 ,\\[2mm]
    \mathcal{L}_V &= - \sum_i \left( \frac{V^{\,i}_{R,{\rm obs}} - V^{\,i}_{R,{\rm th}}}{\sigma^{\,i}_{V}} \right)^2 .
\end{align}
\end{subequations}
Here, $X_i$ and $Y_i$ denote the observed astrometric positions of the S2 star.  
The apparent coordinates depend on the orbital model and on the distance to the Galactic center $r_d$.  
The corresponding theoretical predictions $\hat{X}_i$ and $\hat{Y}_i$ enter via
\begin{subequations}\label{eq.X,Y}
\begin{align}
    X_i &= -\frac{\hat{X}_i}{r_d} + x_0 + v_{x_0}\,(t_{\rm obs}-t_{\rm J2000})\,,\\
    Y_i &= -\frac{\hat{Y}_i}{r_d} + y_0 + v_{y_0}\,(t_{\rm obs}-t_{\rm J2000})\,,
\end{align}
\end{subequations}
where $x_0$, $y_0$, $v_{x_0}$, and $v_{y_0}$ encode the two-dimensional positional offset and the linear drift of the reference frame.  
The quantity $t_{\rm J2000}$ corresponds to the Julian Epoch 2000, and $t_{\rm obs}$ is the time of each measurement \cite{Do:2019txf}.

To compute $\hat{X}_i$ and $\hat{Y}_i$, the three-dimensional orbital coordinates $(x,y,z)$ must be projected onto the plane of the sky.  
The transformation to the apparent coordinates $(\hat{X},\,\hat{Y},\,\hat{Z})$ takes the form
\begin{equation}\label{eq.transform}
    \hat{X} = xA + yB,\qquad
    \hat{Y} = xC + yD,\qquad
    \hat{Z} = xE + yG,
\end{equation}
where the coefficients are given by
\begin{subequations}
\begin{align}
    A &= \sin\Omega\,\cos\gamma + \cos\Omega\,\sin\gamma\,\cos i\,,\\
    B &= -\sin\Omega\,\sin\gamma + \cos\Omega\,\cos\gamma\,\cos i\,,\\
    C &= \cos\Omega\,\cos\gamma - \sin\Omega\,\sin\gamma\,\cos i\,,\\
    D &= -\cos\Omega\,\sin\gamma - \sin\Omega\,\cos\gamma\,\cos i\,,\\
    E &= \sin\gamma\,\sin i\,,\qquad 
    G = \cos\gamma\,\sin i\,,
\end{align}
\end{subequations}
with $\Omega$ the longitude of the ascending node, $i$ the orbital inclination, and $\gamma$ the argument of pericenter.  
The Cartesian coordinates are expressed in terms of the orbital elements as
\begin{equation}\label{eq.trans2}
    x = r\cos\phi,\qquad
    y = r\sin\phi,\qquad
    z = 0.
\end{equation}
The radial velocity of the S2 star is also required in the likelihood evaluation.  
Following Ref. \cite{Do:2019txf,2025PDU....4902022U}, the model prediction reads
\begin{eqnarray}\label{eq.Vr}
    V_R = \xi_D\,\xi_G - 1\,,
\end{eqnarray}
where $\xi_G = 1/\sqrt{-g_{tt}}$ accounts for the gravitational redshift and  
$\xi_D = \sqrt{1-v_{\rm em}^2}/\left( 1 - n\cdot v \right)$ represents the Doppler contribution.  
Here, $v_{\rm em}$ is the orbital velocity of the star at the emission event, and $n$ denotes the unit vector joining the observer and the star.  
The temporal evolution of the modelled radial velocity is illustrated in Fig.~\ref{fig.radial velocity}.
 
\begin{figure*}[htp]
\centering
\includegraphics[width=1\textwidth]{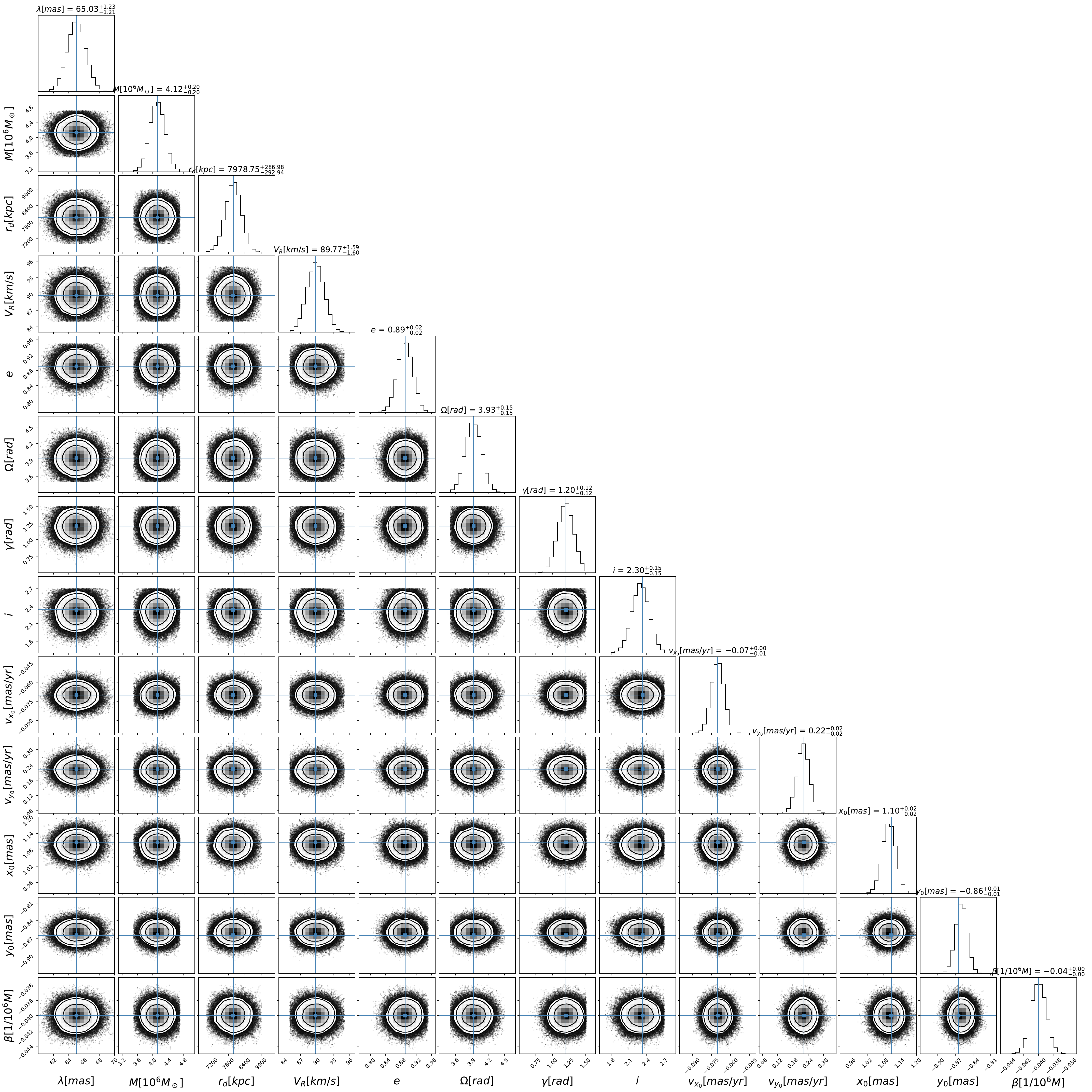}
\caption{Posterior distributions obtained from the MCMC analysis.}
\label{fig.MCMC}
\end{figure*}

For completeness, Fig.~\ref{fig.MCMC} displays the complete set of marginalized likelihoods and joint posterior distributions obtained from our MCMC exploration.  
The diagram summarizes the inferred constraints for the complete parameter set $\{ \lambda,\, M,\, r_d,\, V_R,\, e,\, \gamma,\, i,\, v_{x_0},\, v_{y_0},\, x_0,\, y_0\}$, together with their $1\sigma$ confidence intervals.  
In particular, the contours illustrate the degree of correlation between the orbital elements and the astrometric drift parameters, and confirm that the likelihood peaks sharply around the best–fit values used throughout this work.  
The posterior width of each parameter encodes the relative weight of the astrometric and spectroscopic contributions in Eq.~(\ref{eq.Lik.}), consistent with the structure of the likelihood components defined above.

Furthermore, Table~\ref{Table1} reports the numerical best-fit values extracted from the complete MCMC chains together with their corresponding $1\sigma$ uncertainties. 
\begin{table}[ht!]
    \centering
    \begin{tabular}{|c|c|}
     \hline
       parameters & best-fit values  \\
    \hline
    $\lambda\,[\mathrm{mass}]$   & $65.03^{+1.23}_{-1.21}$  \\
      $M\,[10^6M_\odot]$   & $4.12^{+0.2}_{-0.2}$ \\
      $r_d\,[\mathrm{kpc}]$  & $7978.75^{+286.98}_{-292.94}$ \\
      $V_R\,[\mathrm{km}/\mathrm{s}]$  & $89.77^{+1.59}_{-1.60}$ \\
      $e$  & $0.89^{+0.02}_{-0.02}$ \\
      $\Omega\,[\mathrm{rad}]$  & $3.93^{+0.15}_{-0.15}$ \\
      $\gamma\,[\mathrm{rad}]$  & $1.2^{+0.12}_{-0.12}$ \\
      $i\,[\mathrm{rad}]$  & $2.3^{+0.15}_{-0.15}$ \\
      $v_{x_0}\,[\mathrm{mas}/\mathrm{yr}]$  & $-0.07^{+0.00}_{-0.01}$ \\
      $v_{y_0}\,[\mathrm{mas}/\mathrm{yr}]$  & $0.22^{+0.02}_{-0.02}$ \\
      $x_0\,[\mathrm{mas}]$  & $1.1^{+0.02}_{-0.02}$ \\
      $y_0\,[\mathrm{mas}]$  & $0.86^{+0.01}_{-0.01}$ \\
      $\beta\,[1/10^6M]$  & $-0.04^{+0.001}_{-0.001}$ \\
   \hline
    \end{tabular}
    \caption{Constraint on the best-fit values of the black hole parameters.}
    \label{Table1}
\end{table}
These values summarize the central estimates for the dynamical and astrometric parameters and provide a quantitative reference for the inferred properties of the S2 orbit and the underlying gravitational model.  
The tightness of the intervals reflects the constraining power of the combined astrometric and spectroscopic datasets. At the same time, the consistency between the table and the posterior distributions further confirms the robustness of the parameter extraction procedure implemented here.

An approximate estimate of the charge $q$ of the S2 star can be obtained from the best-fit value of the magnetic coupling parameter $\beta = -0.04\,[1/10^6 M]$. Using
\begin{eqnarray}
    \frac{2\beta m}{B}\cdot\frac{c^{3}}{G} \sim 10^{22}\,\mathrm{C},
\end{eqnarray}
and adopting a magnetic-field strength in the vicinity of Sgr~A* of $B \sim 100\,\mathrm{G}$ \cite{2012A&A...537A..52E}, we obtain the corresponding stellar charge.  
Furthermore, the surface magnetic field of S2 may be estimated following Ref. \cite{2010arXiv1010.1917O} as
\begin{eqnarray}
    B_{0} \sim \frac{\mu_{0} q \Omega}{4\pi R} \sim 10^{3}\,\mathrm{G},
\end{eqnarray}
where $\Omega$ and $R$ denote the rotation rate and radius of the S2 star, respectively.

\section{Conclusions}\label{sec:conclusion}

In this work, we have investigated the motion of charged particles in the Schwarzschild spacetime embedded in a uniform external magnetic field, following Wald’s prescription. By deriving and analyzing the effective potential and the corresponding equations of motion, we identified the key dynamical features governing both circular and noncircular trajectories. The stability analysis of circular motion revealed that the magnetic coupling parameter $\beta$ plays a decisive role in modifying the location of the ISCO, as well as the associated constants of motion.

The study further explored the astrophysical implications of these results in two important observational contexts: the orbital motion of the S2 star and the near-horizon dynamics of bright hotspots detected around Sgr~A*. By constraining the allowed values of $\beta$ using GRAVITY observations, we demonstrated that small magnetic couplings can produce measurable deviations in the orbital period, offering a potential explanation for magnetically induced perturbations in stellar and plasma trajectories near the galactic center.

In addition to these analytical considerations, we conducted a full MCMC analysis to determine the best-fit model parameter values from the combined astrometric and spectroscopic data. The resulting posterior distributions and likelihood contours confirmed the internal consistency of the framework and yielded tight $1\sigma$ constraints for $\{\lambda,\, M,\, r_d,\, V_R,\, e,\, \Omega,\, \gamma,\, i,\, v_{x_0},\, v_{y_0},\, x_0,\, y_0,\, \beta\}$. The associated parameter table provides a quantitative summary of these results, showing that the inferred values of $\beta$ remain in the weak-coupling regime and are fully compatible with the weak magnetic-induced deviations predicted by our dynamical model. This numerical analysis, therefore, reinforces the observational viability of the scenario studied here and anchors the theoretical predictions in a statistically robust inference procedure.

Using the LP taxonomy, we classified relativistic bound trajectories and provided numerical examples illustrating how the Lorentz factor $\beta$ and eccentricity $e$ influence orbital morphology. The family of periodic and quasiperiodic solutions obtained highlights how weak magnetization can give rise to distinctive zoom–whirl structures and nontrivial precession patterns, potentially observable through high-precision astrometric monitoring.

A notable outcome of our analysis is the extremely small inferred charge on the S2 star via the magnetic coupling parameter $\beta$. Although this value appears surprisingly low, it is entirely consistent with astrophysical expectations: a massive object such as S2 cannot retain a significant net charge, since the surrounding plasma and accretion of opposite charges would rapidly neutralize any excess. Moreover, the charge entering the dynamics should be understood as an \emph{effective} charge--to--mass coupling describing the interaction between the stellar plasma and the external magnetic field. Even a minute charge can lead to observable dynamical effects when combined with strong magnetization and relativistic orbital velocities near Sgr~A*. Thus, the smallness of the inferred $q$ reflects the natural physical limits on stellar charging and the high sensitivity of the orbital dynamics to weak magnetization, and it remains fully compatible with the scenario explored in this work.

To summarize, our results underline the importance of magnetic effects in shaping the relativistic dynamics of charged particles near black holes. Future extensions of this work may include rotating (Kerr) geometries, plasma backreaction, and radiative energy losses, thereby offering a more complete description of magnetized accretion and emission phenomena in strong-gravity environments.

\section*{Acknowledgements}
The work of M.F. has been supported by Universidad Central de Chile through the project No. PDUCEN20240008.

\appendix

\section{The coefficients in Eqs. \eqref{eq:angular} and \eqref{eq:energy}}\label{app:A}

These coefficient are as follows:
\begin{eqnarray} \mathcal{A}=\lambda ^2 M\beta \left(1-e^2\right)^2, \end{eqnarray}
\begin{eqnarray} 
\mathcal{B}&=&\Biggl\{(1-e^2)^2 \lambda ^2 \Biggl[\beta ^2 \lambda ^4-(1-e^2) M^2 (4 \beta ^2 \lambda ^2+e^4+2 e^2\nonumber\\
&&-3)+\lambda M (-4 \beta ^2 \lambda ^2+e^4-2 e^2+1)\Biggr]\Biggr\}^{1/2},
\end{eqnarray} 
\begin{eqnarray} 
\Psi=(1-e^2)^2 \Big[(e^2+3) M-\lambda \Big], 
\end{eqnarray} 
\begin{eqnarray} 
\mathcal{C}=4 (1-e^2) M^2-\lambda ^2+4 \lambda M, 
\end{eqnarray} 
\begin{eqnarray} 
\mathcal{D}=\lambda\left(-2 \beta ^2 \lambda ^2-e^4-2 \beta ^2 e^2 \lambda ^2+2 e^2\right), 
\end{eqnarray} 
\begin{eqnarray} 
\mathcal{G}&=&2 \beta \Bigg\{(1-e^2)^2 \lambda ^2 \Biggl[\beta ^2 \lambda ^4-(1-e^2) M^2 (4 \beta ^2 \lambda ^2+e^4+2 e^2\nonumber\\
&&-3)+\lambda M (-4 \beta ^2 \lambda ^2+e^4-2 e^2+1)\Biggr]\Bigg\}^{1/2}, 
\end{eqnarray} 
\begin{eqnarray} 
\mathcal{H}=M\Biggl[4 \beta ^2 \lambda ^2+e^6+e^4+e^2 (12 \beta ^2 \lambda ^2-5)+3\Biggr], 
\end{eqnarray} 
\begin{eqnarray} 
\Xi=(e^2+3) M-\lambda.
\end{eqnarray}

\bibliographystyle{apsrev4-1}  
\bibliography{Ref.bib}
\end{document}